\documentclass{ws-ijmpe}

\begin{document}

\markboth{Masayasu Harada, Yukio Nemoto, and Shunji Yoshimoto}
{Quark spectrum above the critical temperature from Schwinger-Dyson
equation}

%%%%%%%%%%%%%%%%%%%%% Publisher's Area please ignore %%%%%%%%%%%%%%%
\catchline{}{}{}{}{}
%%%%%%%%%%%%%%%%%%%%%%%%%%%%%%%%%%%%%%%%%%%%%%%%%%%%%%%%%%%%%%%%%%%%

\title{QUARK SPECTRUM ABOVE THE CRITICAL TEMPERATURE FROM SCHWINGER-DYSON
  EQUATION}

\author{\footnotesize Masayasu Harada, Yukio Nemoto, and Shunji Yoshimoto
\footnote{Poster presented by S. Y. at Quark Matter 2006}}

\address{Department of Physics, Nagoya University, Nagoya,
Aichi 464-8602, Japan\\
yoshimoto@hken.phys.nagoya-u.ac.jp}

%\author{SECOND AUTHOR}
%
%\address{Group, Laboratory, Address\\
%City, State ZIP/Zone, Country\\
%second\_author@group.com}

\maketitle

\begin{history}
%\received{(received date)}
%\revised{(revised date)}
%\accepted{(Day Month Year)}
%\comby{(xxxxxxxxxx)}
\end{history}

\begin{abstract}

We investigate a spectrum of a fermion, which we call a quark,
above the critical temperature of
the chiral phase transition in a gauge theory using the Schwinger-Dyson (SD)
equation.
The SD equation enables us to study the spectrum over a wide range of
the gauge coupling.
It is shown that the quark spectrum has two sharp peaks which correspond to
the normal quasi-quark and the plasmino and is
 consistent with that obtained in the
hard thermal loop approximation in the weak coupling region, 
while it has also
two peaks but with smaller thermal masses and broader widths in the strong
coupling region.
Temperature-dependence of the quark spectrum is also discussed.
\end{abstract}

\section{Introduction}

Recent experimental researches performed at the Relativistic Heavy Ion 
Collider (RHIC) reveal unexpected findings\cite{Arsene:2004fa}.
Among them,
collective flow of the created matter behaves like a perfect fluid,
which suggests that the quark-gluon plasma (QGP) near the phase
transition is a strongly interacting system.
It seems to be consistent with the recent studies in Lattice QCD,
showing that the lowest charmonium states survive at temperature ($T$)
 higher than the critical temperature ($T_c$) of 
deconfinement\cite{Asakawa:2003re}.
Theoretically, hadronic states in QGP were already predicted 
many years ago on the basis of symmetry arguments of the chiral phase 
transition\cite{Hatsuda:1985eb}.
There are also many recent studies on possible hadronic bound states in QGP
motivated by the RHIC experiments\cite{Shuryak:2004cy,Brown:2003km}.

The study of quarks and gluons as well as the hadronic states is also
important, because they are explicitly deconfined degrees of freedom in QGP.
Rapid change of the energy density around $T_c$ obtained in Lattice QCD
suggests that quarks and gluons actually come into play in thermodynamics.
However, particles in medium have, in general, different spectra from 
those in vacuum,
and thus it is quite nontrivial whether quarks and gluons keep the
quasi-particle picture in the strongly coupled matter like QGP near $T_c$.
It is shown in Ref.~\refcite{Schaefer:1998wd} that 
the quark spectrum near $T_c$ has two massive modes
taking the gluon condensate into account.
In Ref.~\refcite{Petreczky:2001yp}, 
dispersion relations of the quark and the gluon above $T_c$ of
the deconfinement transition are obtained in quenched Lattice QCD with the 
maximum entropy method.
They do not find any pure collective modes like the plasmino, and
the thermal (pole) masses of the quasi-quark and the quasi-gluon are
 higher than $T$ near $T_c$.
In Ref.~\refcite{Mannarelli:2005pz}, 
using a Brueckner-type many-body scheme together with data from
the heavy quark potential in Lattice QCD, the thermal mass of quark
is obtained as $m_q\sim100$ MeV for $T=1$-$2 T_c$.
In Ref.~\refcite{Kitazawa:2005mp}, it is shown that 
the quark spectrum near $T_c$ of the chiral phase transition has three peaks
through a coupling with fluctuations of the chiral condensate.

In this paper, we investigate the quark spectrum near but above $T_c$ of 
the chiral phase transition in a {\it gauge theory}, where we call a fermion
a ``quark''.
In a weak coupling region at finite $T$, the HTL
resummation is established, and the quark spectrum is well understood at
the leading order\cite{Klimov:1981ka}.
In this study, we compute the quark spectrum in the chiral restored phase
over a wide range of the coupling constant, especially in {\it the strong
coupling region}, using the Schwinger-Dyson (SD) equation.
The SD approach has been employed in analyses of the strong coupling 
gauge theory in vacuum\cite{kugo} and in the determination of the 
phase structure of the chiral and color superconducting
transitions at finite $T$ and/or 
density\cite{Harada:1998zq,Ikeda:2001vc}.

The SD equation used in this paper includes HTL of the quark self-energy
at the leading oder, and thus reproduces the quark spectrum of HTL in the
weak coupling region.
In the strong coupling region, on the other hand, the SD equation incorporates
nonperturbative corrections as an infinite summation of a certain kind of 
diagrams.
In addition, 
it is one of advantageous points that the SD equation respects the chiral
symmetry and describes the dynamical breaking of it, while
it is a heavy task to respect the chiral symmetry on the lattice.
Furthermore, it is straightforward to extend our formulation to finite density.

The contents of this paper are as follows.
In Sec. 2, we formulate the SD equation in the imaginary time formalism
with the ladder approximation. 
Then we perform an analytic continuation of the quark propagator 
to the real-time axis.
We perform it numerically with a method of an integral equation\cite{ac}.
Using the retarded quark propagator obtained in this way,
we investigate the quark spectrum above $T_c$, focusing on dependences
on the gauge coupling and $T$, in Sec. 3.

 %%%%%%%%%%%%%%%%%%%%%%%%%%%%%%%%%%%%%%%%%%%%%%%%%%%%%%%%%%%%%%%%%%%%
\section{Schwinger-Dyson Equation}
\label{sec:SDE}
  
 The SD equation in the ladder approximation is given by
\begin{eqnarray}
  \lefteqn{ \mathcal{S}^{-1}(i\omega_n,\vec{p})-
  \mathcal{S}^{-1}_{\rm{free}}(i\omega_n,\vec{p})} \nonumber \\ &=& 
  \frac{4}{3} \, T\sum_{m=-\infty}^{\infty}\int\frac{d^3k}{(2\pi)^3}
  g^2\gamma_{\mu}\mathcal{S}(i\omega_m,\vec{k})
  \gamma_{\nu}\mathcal{D}^{\mu\nu}_{\rm{free}}
  (i\omega_n-i\omega_m,\vec{p}-\vec{k}),
  \label{eq:ladderSDE}
\end{eqnarray}
where $\mathcal{S}_{\rm{free}}$ and $\mathcal{D}^{\mu\nu}_{\rm{free}}$are the
free massless fermion and free gauge boson
propagator, respectively.
We employ the imaginary time formalism and
$\omega_n=(2n+1)\pi T$ is Matsubara frequency for fermion. 
$\mathcal{S}$ stands for the full fermion propagator which is written as
\begin{eqnarray}
\mathcal{S}(i\omega_n,\vec{p})=\frac{1}{C(i\omega_n,p)i\omega_n\gamma_0-A(i\omega_n,p)\vec{p}\cdot\vec{\gamma}-B(i\omega_n,p)},
\label{eq:Sm}
\end{eqnarray}
from the rotational invariance in space and the parity invariance.
The gauge coupling $g$ is not running in this study, and thus an
ultraviolet cutoff comes in.
We scale all the dimension-full quantities by a three-momentum cutoff.

At zero $T$, the Landau gauge is often chosen in the SD equation 
with the ladder approximation, because it satisfies the Ward-Takahashi 
identity in the case of QED. 
At finite $T$, on the other hand, any constant gauge-fixing
 parameter including the
Landau gauge does not satisfy the Ward-Takahashi 
identity.\cite{Ikeda:2001vc},
We do not pursue this issue here and adopt the Feynman gauge in the following,
which makes the analytic continuation simple.

To study the quark spectrum,
we need the analytic continuation of the solutions of Eq.(\ref{eq:ladderSDE})
obtained on the imaginary axis in the complex energy plane to the real
axis.
In this study, following a method proposed by Ref.~\refcite{ac}, 
the analytic continuation
is done numerically.
The retarded quark Green function $S^R(p_0,\vec{p})$
is then obtained by solving a following
integral equation:
\begin{eqnarray}
\lefteqn{S_{R}^{-1}(p_0,\vec{p})-S_{R\,\,{\rm free}}^{-1}(p_0,\vec{p}) }
\nonumber \\
&=& \frac{4}{3} \int\frac{d^3\vec{k}}{(2\pi)^3}\int_{-\infty}^{\infty} dz g^2
T\sum_{m=-\infty}^{\infty}\gamma_{\mu}
\left[\frac{\mathcal{S}(i\omega_m,\vec{k})}{p_0-z-i\omega_m}\right] \gamma_{\nu}
\rho_{B}^{\mu\nu}(z,\vec{p}-\vec{k})
\nonumber \\
& & -\frac{4}{3} \int\frac{d^3\vec{k}}{(2\pi)^3}
\int_{-\infty}^{\infty}dz 
g^2 \gamma_{\mu}
S_R(p_0-z,\vec{k})\gamma_{\nu} \rho_{B}^{\mu\nu}(z,\vec{p}-\vec{k})\frac{1}{2}
\left[{\rm tanh}\frac{p_0-z}{2T}+{\rm
coth}\frac{z}{2T}\right]\nonumber \\
\label{eq:acSDeq}
\end{eqnarray}
where $\rho_B^{\mu\nu}$ stands for the spectral function for the gauge boson,
and $\mathcal{S}$ is the solution of Eq.(\ref{eq:ladderSDE}).
This method is known as a reliable way of the analytic continuation in
condensed matter physics,\cite{ac} and suitable for the SD approach because the
SD equation Eq.(\ref{eq:ladderSDE}) is also an integral equation
which is similar to Eq.(\ref{eq:acSDeq}).

  %%%%%%%%%%%%%%%%%%%%%%%%%%%%%%%%%%%%%%%%%%%%%%%%%%%%%%%%%%%%%%%%%%%%
\section{Quark spectrum}
\label{sec:spf}

The quark spectral function is given by 
\begin{eqnarray}
  \rho_{\pm}(p_0,p) = -\frac{1}{\pi}{\rm Im} 
  {\rm Tr}\left[\frac{\gamma_0\pm \vec{\gamma}\cdot\hat{\vec{p}}}{2} S_R \right] 
\end{eqnarray}
where $\rho_+(\rho_-)$ denotes the quark (anti-quark) spectrum 
at zero $T$, and in addition, the anti-plasmino (plasmino) spectrum, 
if they exist at finite $T$.
We show only $\rho_+$ in the following, because there is a relation of
$\rho_{-}(p_0,p)=\rho_{+}(-p_0,p)$.

\begin{figure}[t]
  \begin{tabular}{cc}
   \begin{minipage}{0.5\hsize}
      \begin{center}
	\includegraphics[keepaspectratio,height=4cm]{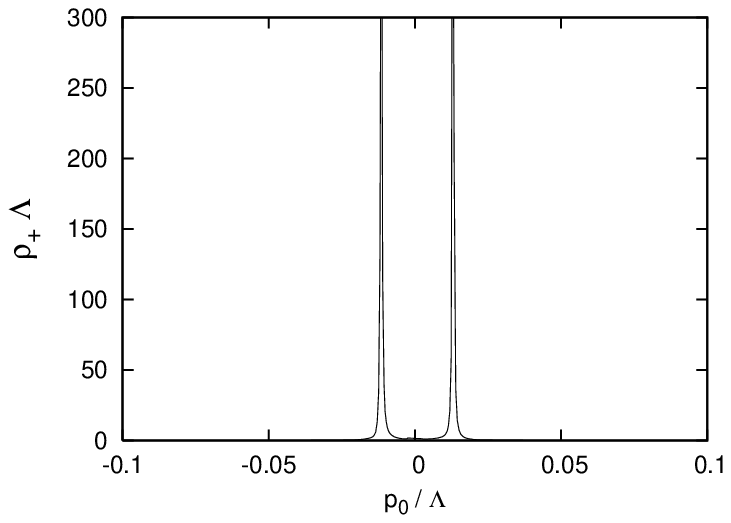}
	\caption{The spectral function $\rho_+$ for $p = 0$, 
	  $g=0.1$, and $\rm{T}/\Lambda=0.3216$.}
	\label{fig:sp-g01}
      \end{center}
   \end{minipage}
   \begin{minipage}{0.5\hsize}
     \begin{center}
       \includegraphics[keepaspectratio,height=4cm]{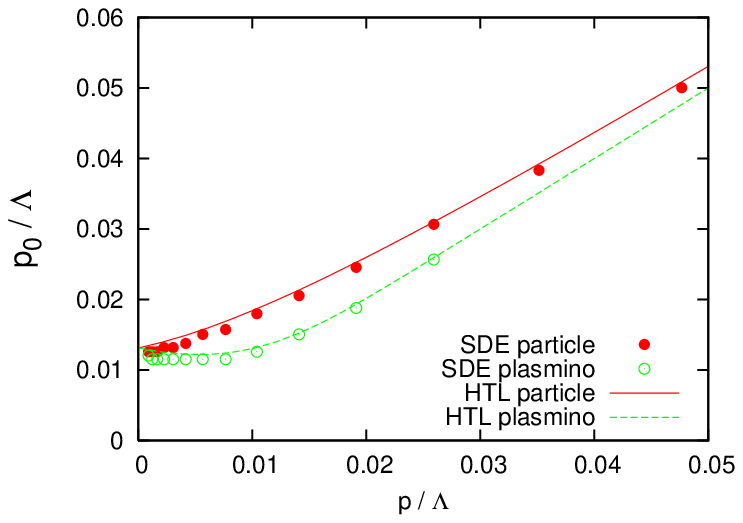}
       \caption{The dispersion relation for the normal quasi-particle and 
      the plasmino for $g=0.1$ and $\rm{T}/\Lambda=0.3216$.}
       \label{fig:dis-g01}
     \end{center}
   \end{minipage}
  \end{tabular}
\end{figure}

We first investigate the quark spectrum in the weak coupling region.
The spectral function $\rho_+$ for $p=0$, $g=0.1$, and
$T/\Lambda = 0.3216$ is shown in Fig.\,\ref{fig:sp-g01}. 
We see that $\rho_{+}$ has two sharp peaks:
One is in the positive energy region and corresponds to the normal
quasi-quark, and the other in the negative region to the 
anti-plasmino.
From the relation between $\rho_+$ and $\rho_-$ mentioned above,
the quasi-antiquark and the plasmino also appear.
Positions of these peaks with the positive energy for finite 
momentum are shown in
Fig.\ref{fig:dis-g01}, which give approximately 
dispersion relations of the modes.
We see that the dispersion relations obtained from the SD equation 
are in good agreement with those in the HTL approximation which is
valid in the weak coupling limit.
With the narrow widths of the peaks,
this means that the spectrum from the SD equation reproduce that
in the HTL approximation correctly in the weak coupling region.

\begin{figure}[ht]
  \begin{tabular}{ccc}
    \begin{minipage}{1.0\hsize}
  \begin{minipage}{0.3\hsize}
      \begin{center}
	\includegraphics[keepaspectratio,height=3.5cm]{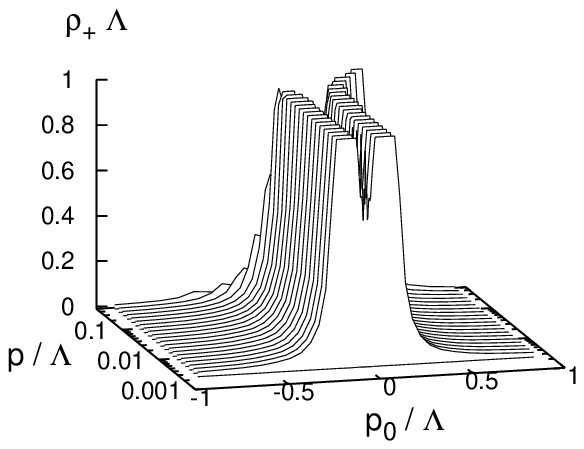}
      \end{center}
    \end{minipage}
   \begin{minipage}{0.3\hsize}
      \begin{center}
	\includegraphics[keepaspectratio,height=3.5cm]{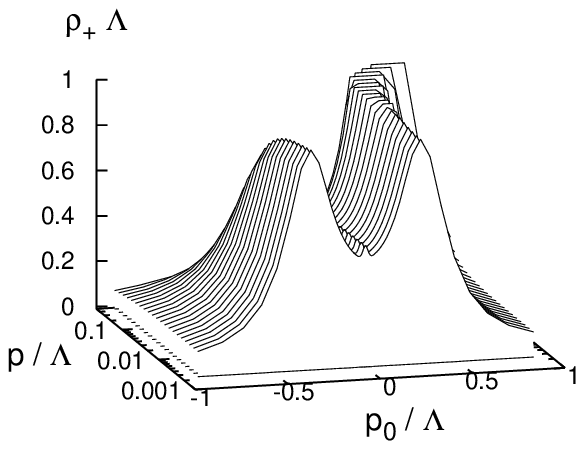}
      \end{center}
    \end{minipage}
   \begin{minipage}{0.3\hsize}
      \begin{center}
	\includegraphics[keepaspectratio,height=3.5cm]{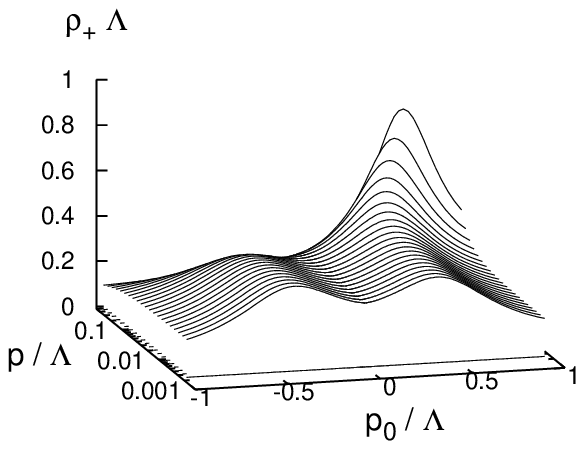}
      \end{center}
    \end{minipage}\\
      \caption{The spectral functions $\rho_+$ for $T/\Lambda=0.3216$ and 
       $g=1$(left), $g=3$(middle), and $g=6$(right).}
      \label{fig:sp-g-depnd}
    \end{minipage}
  \end{tabular}
\end{figure}

For the strong coupling, we plot the coupling dependence of 
the spectral functions $\rho_+$
for $T/\Lambda=0.3216$\ in Fig.\,\ref{fig:sp-g-depnd}.
There also exist two peaks but with broad widths, showing
that the normal quasi-quark and the plasmino appear even in the strong
coupling region.
We see that as the coupling is larger, the widths are broader.
It could be understood from the fact that probability of gluon emission
and absorption from a quark increases with the coupling.
In fact, these peaks obtained at the one loop order without 
the HTL approximation are already broad in the strong coupling region. 
In an extremely strong coupling, e.g. $g=6$, the peaks are so broad
for low momentum that the quasi-particle picture is no longer valid.

As momentum is higher, the peak of the plasmino is smaller, which
is consistent with the momentum-dependence of the strength of the 
plasmino in the HTL approximation.
On the other hand, the peak of the normal quasi-quark is sharper for
higher momentum.
This is because thermal effects get smaller as momentum
increases, and thus the spectrum approaches that at zero $T$.

\begin{figure}[htbp]
  \begin{tabular}{ccc}
   \begin{minipage}{1.0\hsize}
      \begin{center}
	\includegraphics[keepaspectratio,height=4cm]{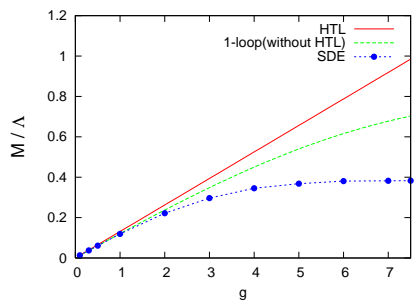}
      \end{center}
      \caption{$g$-dependence of the thermal mass for $\rm{T}/\Lambda=0.3216$. 
      ``1-loop(without HTL)'' (``HTL'') denotes the thermal mass at one loop
      without (with) the HTL approximation.}
      \label{fig:t03216-Mth-g}
    \end{minipage}
  \end{tabular}
\end{figure}

Positions of the peaks for $p=0$ give approximately the thermal masses
of the modes.
In Fig.\,\ref{fig:t03216-Mth-g} we plot the coupling dependence of the
thermal mass.
We note that the thermal masses of the normal quasi-quark and the plasmino
are same in the chiral limit at zero density.
In Fig.\,\ref{fig:t03216-Mth-g}, we also plot the thermal mass
obtained at the one-loop order with and without the HTL approximation.
This shows that the thermal mass in the weak coupling
is proportional to the coupling $g$ and 
coincides with the one obtained in the HTL approximation. 
In the strong coupling region, on the other hand,
the thermal mass is almost independent of the coupling $g$ and
expressed as $M \sim T$. 
We have checked that this property is independent of $T$.
As a result, the thermal mass is smaller than the one obtained in the HTL
approximation and at the one-loop order without the HTL approximation. 

Finally, we mention the $T$-dependence of the quark spectrum
in the strong coupling region.
We see that the thermal mass is proportional to $T$, which means that
it is determined only by the infrared
dynamics characterized by $T$.
The widths of the peaks are broader as $T$ increases, because
damping effects resulting from the interaction with thermal particles
are more significant for higher $T$.

This work is supported in part by the 21st Century COE Program
at Nagoya University, 
the JSPS Grand-in-Aid for Scientific Research \#18740140 (Y.N.),
the Daiko Foundation \#9099 (M.H.)
and the JSPS Grant-in-Aid for Scientific Research (c) (2) \#16540241
(M.H.).

\end{document}